\documentclass[useAMS, usenatbib, usegraphicx]{mn2e}
\pdfoutput=1

\usepackage[fleqn]{amsmath} 
\usepackage{amsfonts} 
\usepackage{amstext}
\usepackage{amssymb} 
\usepackage[colorlinks=true, citecolor=blue, linkcolor=blue, breaklinks=true]{hyperref}

\newcommand{\hi } {{\rm H}\,{\small\rm I} \,}
\newcommand{\hii } {{\rm H}\,{\small\rm II} \,}
\newcommand{\msun}{M$_{\odot}$}
\newcommand{\ltsimeq}{\la}

\title[The link between mass distribution and starbursts in dwarf galaxies]{The link between mass distribution and starbursts in dwarf galaxies\thanks{Based on observations made with the NASA/ESA Hubble Space Telescope, and obtained from the Hubble Legacy Archive, which is a collaboration between the Space Telescope Science Institute (STScI/NASA), the Space Telescope European Coordinating Facility (ST-ECF/ESA) and the Canadian Astronomy Data Centre (CADC/NRC/CSA).}}
\author[K. McQuinn et al.]
{Kristen~B.~W. McQuinn$^1$\thanks{E-mail:kmcquinn@astro.umn.edu},
Federico Lelli,$^2$
Evan D.~Skillman,$^1$
\newauthor
Andrew E.~Dolphin,$^3$
Stacy S.~McGaugh,$^2$
Benjamin F.~Williams$^4$\\
$^1$Minnesota Institute for Astrophysics, School of Physics and
Astronomy, 116 Church Street, S.E.,\\
University of Minnesota, Minneapolis, MN 55455\\
$^2$Department of Astronomy, Case Western Reserve University, Cleveland, OH 44106\\
$^3$Raytheon Company, 1151 E. Hermans Road, Tucson, AZ 85756\\
$^4$Department of Astronomy, Box 351580, University of Washington, Seattle, WA 98195}

\begin{document}

\date{}
\maketitle
 
\begin{abstract}
Recent studies have shown that starburst dwarf galaxies have steeply rising rotation curves in their inner parts, pointing to a close link between the intense star formation and a centrally concentrated mass distribution (baryons and dark matter). More quiescent dwarf irregulars typically have slowly rising rotation curves, although some ``compact'' irregulars with steep, inner rotation curves exist. We analyze archival Hubble Space Telescope images of two nearby ``compact'' irregular galaxies (NGC~4190 and NGC~5204), which were selected solely on the basis of their dynamical properties and their proximity. We derive their recent star-formation histories by fitting color-magnitude diagrams of resolved stellar populations, and find that the star-formation properties of both galaxies are consistent with those of known starburst dwarfs. Despite the small sample, this strongly reinforces the notion that the starburst activity is closely related to the inner shape of the potential well. 
\end{abstract}


\section{Connecting Star-Formation Characteristics and Mass Distributions}

Despite numerous and extensive investigations, the mechanisms that trigger, fuel, and quench starbursts in galaxies remain poorly understood. High-mass starburst galaxies, such as luminous and ultra luminous infrared galaxies (LIRGs and ULIRGs), are  typically associated with major mergers and tidal interactions \citep[e.g.,][]{Sanders1996}, suggesting that external, violent events are needed to trigger the intense star formation. In contrast, circumnuclear starbursts in isolated spiral galaxies seem to be associated with stellar bars \citep[e.g.,][]{Kennicutt1998, Kormendy2004}, pointing to internal, secular processes. The situation is much less clear for low-mass starburst galaxies, such as blue compact dwarfs \citep[BCDs;][]{GilDePaz2003} and \hii galaxies \citep[e.g.,][]{Taylor1995}, which generally lack stellar bars and spiral arms. Several authors \citep[e.g.,][]{Ekta2010a, LopezSanchez2010a, Lelli2014c} argue that external mechanisms may be important in triggering starbursts in at least some of these systems, but it is generally difficult to distinguish between dwarf-dwarf mergers, past interactions between unbound, neighboring galaxies, or cold gas accretion from the intergalactic medium. Detailed studies of the stellar and dynamical properties of starburst dwarfs, however, can provide indirect clues on the starburst trigger.

\begin{figure*}
\includegraphics[width=0.49\linewidth]{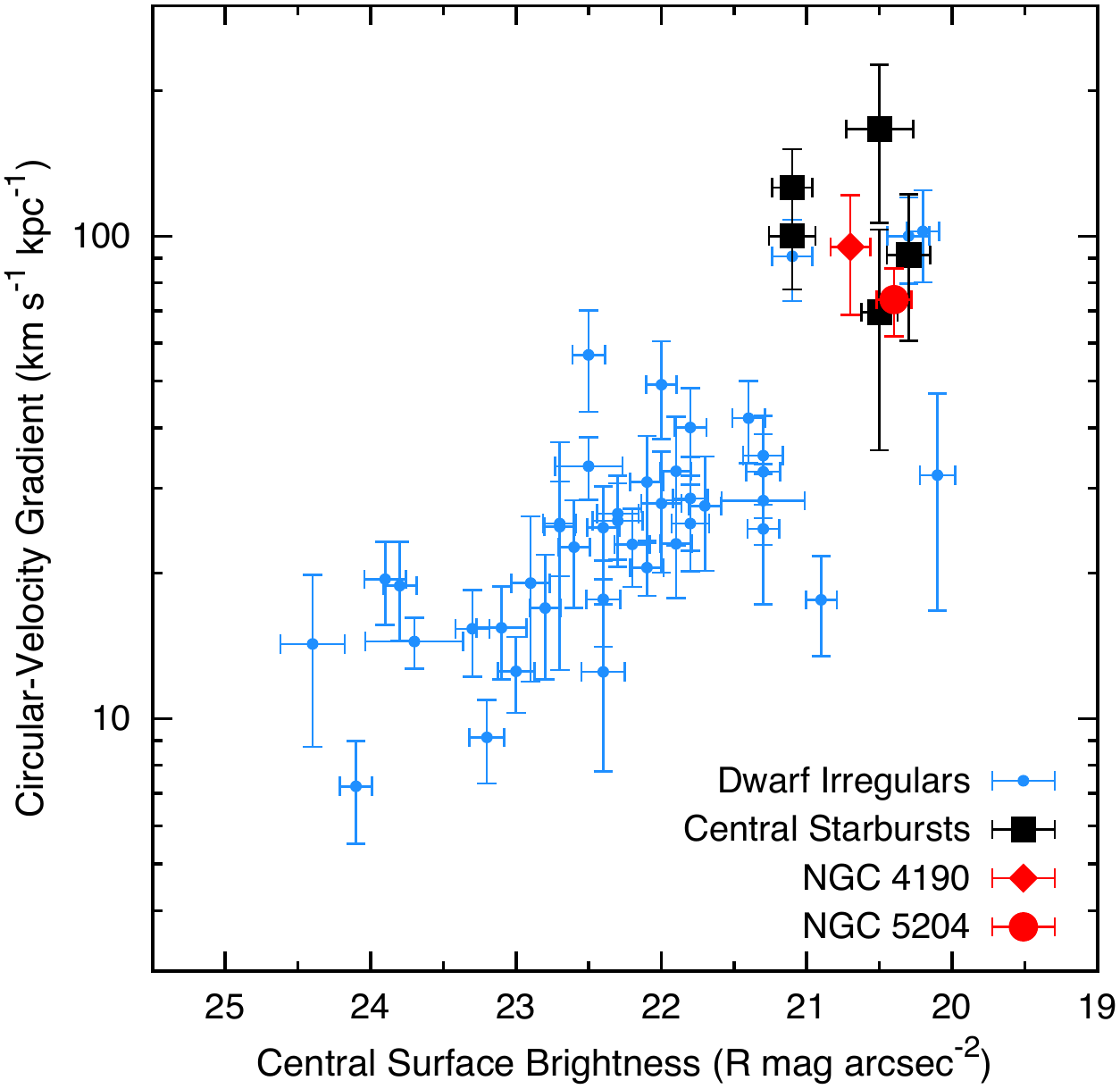}
\includegraphics[width=0.49\linewidth]{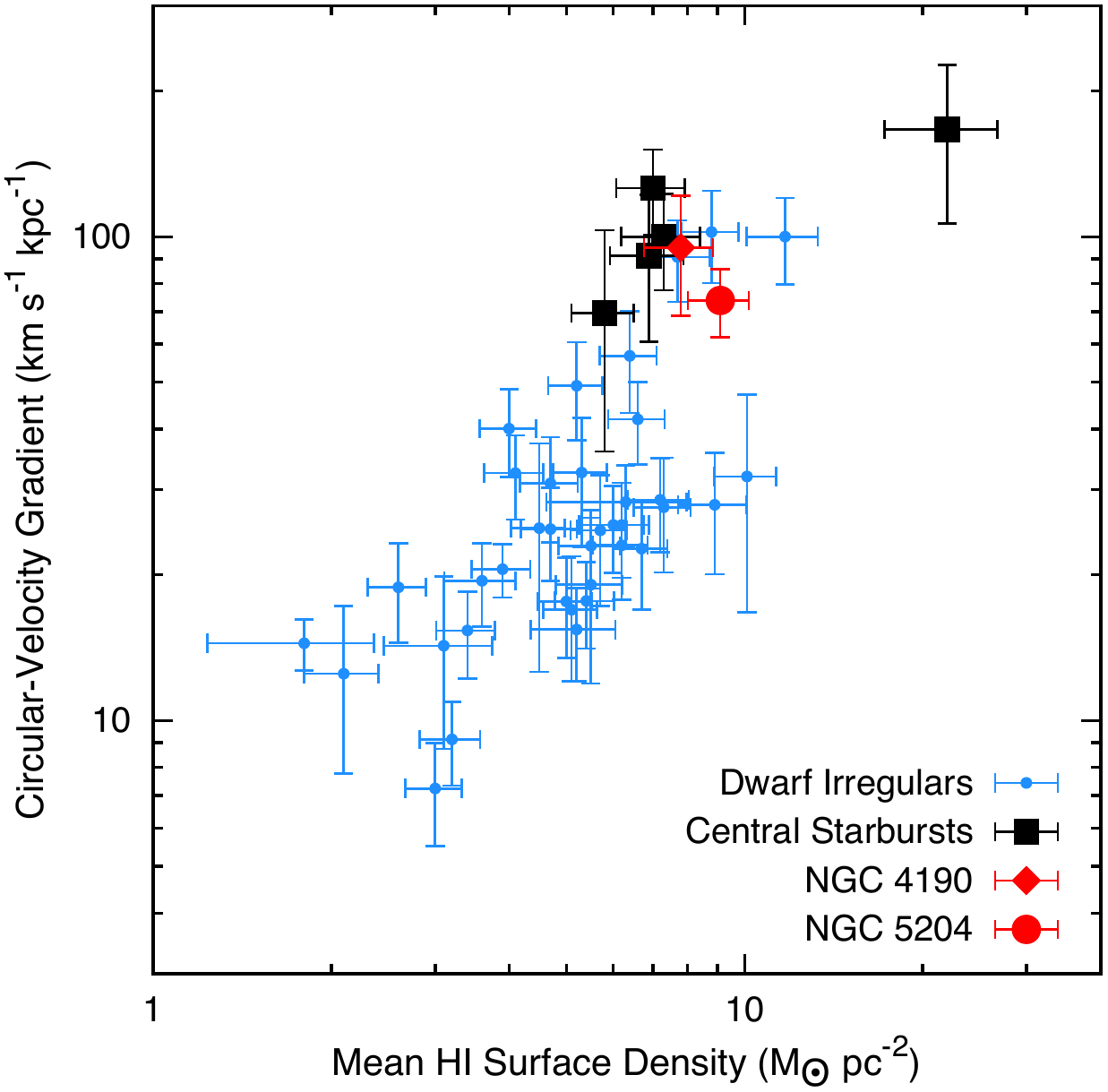}
\caption{\textit{Left:} Circular-velocity gradient ($V_{R_{\rm d}}/R_{\rm d}$) and central surface brightness for dwarf galaxies. \textit{Right:} $V_{R_{\rm d}}/R_{\rm d}$ versus the mean \hi surface density within the optical radius. Both figures are adapted from \citet{Lelli2014a}. For clarity, we omit galaxies with off-centered or diffuse starbursts, such as cometary BCDs. Galaxies with a previously-identified central starburst (black symbols) lie in the upper right portion of both relations. A few dIrrs (cyan symbols) are found in the same region: two of them, NGC~4190 and NGC~5204 (red symbols), have been imaged with the HST and constitute the focus of this paper.}
\label{fig:gradient}
\end{figure*}

Regarding stellar properties, \citet{McQuinn2010a, McQuinn2010b} used Hubble Space Telescope (HST) observations of 20 active starburst and post-starburst dwarfs in the Local Volume to derive accurate star-formation histories (SFHs) from fitting color-magnitude diagrams (CMDs) of resolved stellar populations. They defined a ``starburst'' as the period when the star-formation rate (SFR) is enhanced by at least a factor of 2 with respect to the historical average SFR in the galaxy, similar to previous prescriptions \citep[i.e., birthrate parameter $b = \rm{SFR/\overline{SFR}}\gtrsim2$;][]{Scalo1986, Kennicutt2005}, and found burst durations of the order of a few 100 Myr. Moreover, \citet{McQuinn2012a} studied the spatial distribution of the resolved stellar populations and found that the starburst component can cover a large fraction of the underlying host galaxy. Similar results have also been found in detailed studies of individual cases \citep[e.g.,][]{Annibali2003, Annibali2013}. It is clear, therefore, that starbursts in dwarf galaxies are not simply stochastic, small-scale fluctuations in the SFH, but they are major events that could, in principle, dramatically affect its structure and evolution.

Regarding internal dynamics, \citet{Lelli2012a, Lelli2012b, Lelli2014a} found that the inner rotation curves of starburst dwarfs rise more steeply than those of typical dwarf irregulars \citep[dIrrs; see also][]{Meurer1998, vanZee2001}, indicating that starburst galaxies have a higher central dynamical mass density (gas, stars, and dark matter). This points to a close connection between the mass distribution in a galaxy and its star formation activity. In particular, \citet{Lelli2014a} considered a sample of 60 low-mass galaxies with high quality rotation curves (including dIrrs, spheroidals, and starburst dwarfs) and estimated their inner circular-velocity gradient as $V_{R_{\rm d}}/R_{\rm d}$, where $R_{\rm d}$ is the exponential scale length. As shown in Figure~\ref{fig:gradient}, they found that $V_{R_{\rm d}}/R_{\rm d}$ correlates with the central surface brightness and the disc-averaged \hi surface density, as well as the disc-averaged SFR surface density. Galaxies with central starbursts, such as BCDs, are systematically found to have high values of $V_{R_{\rm d}}/R_{\rm d}$, but there are also several ``compact'' dIrrs with similarly high values of $V_{R_{\rm d}}/R_{\rm d}$. The nature of these compact dIrrs is unclear: they may be the progenitors/descendants of BCDs caught in a quiescent phase, or they may be undergoing a starburst at the present time.

We identified a sample of seven ``compact'' dIrrs by selecting galaxies in the top-right end of the $V_{R_{\rm d}}/R_{\rm d}-\mu_0$ relation in Figure~\ref{fig:gradient} ($V_{R_{\rm d}}/R_{\rm d} \gtrsim 50$ km s$^{-1}$ kpc$^{-1}$, $\mu_0 \ltsimeq 22$ mag arcsec$^{-2}$). Two of these (NGC~4190 and NGC~5204) have archival HST observations that meet the criteria of \citet{McQuinn2010a} for deriving SFHs. First, the galaxies were close enough such that their stellar populations were resolved by HST imaging instruments. Second, both V and I band images of the galaxy were available in the HST archive. Third, the I band observations had a minimum photometric depth of $\sim$2 mag below the tip of the red giant branch; a requirement for accurately constraining the recent SFH of a galaxy \citep{Dolphin2002, Dohm-Palmer2002}. 

In this paper, we drive the SFHS of NGC~4190 and NGC~5204 using a CMD fitting technique. We find that both galaxies have star-formation properties that are consistent with those of known starbursts. This strongly reinforces the idea that a high central mass density (baryons and dark matter) is a key property of starburst galaxies, which must be closely linked to the mechanism that triggered the intense star-formation.

\begin{figure*}
\includegraphics[width=0.5\textwidth]{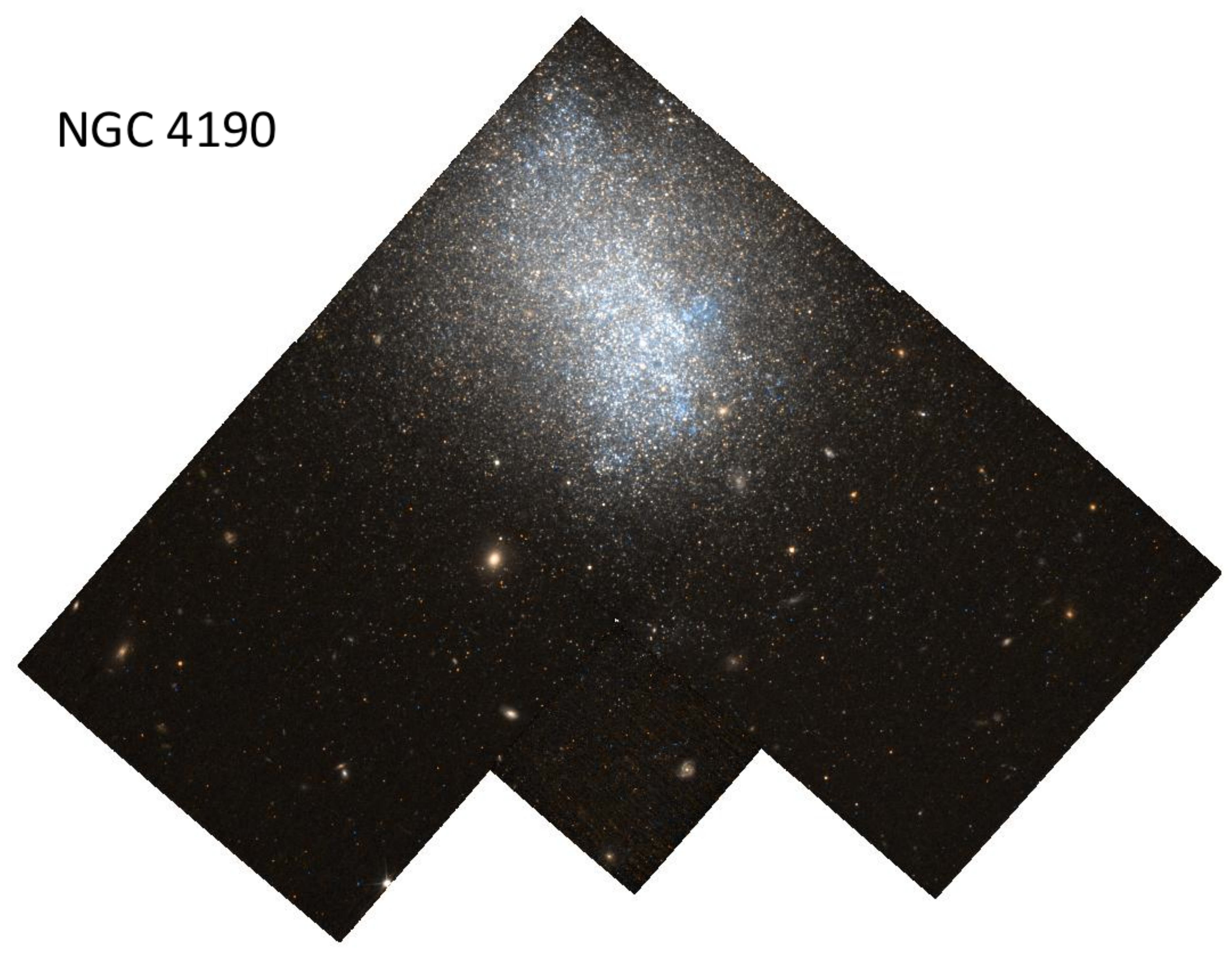}
\includegraphics[width=0.42\textwidth]{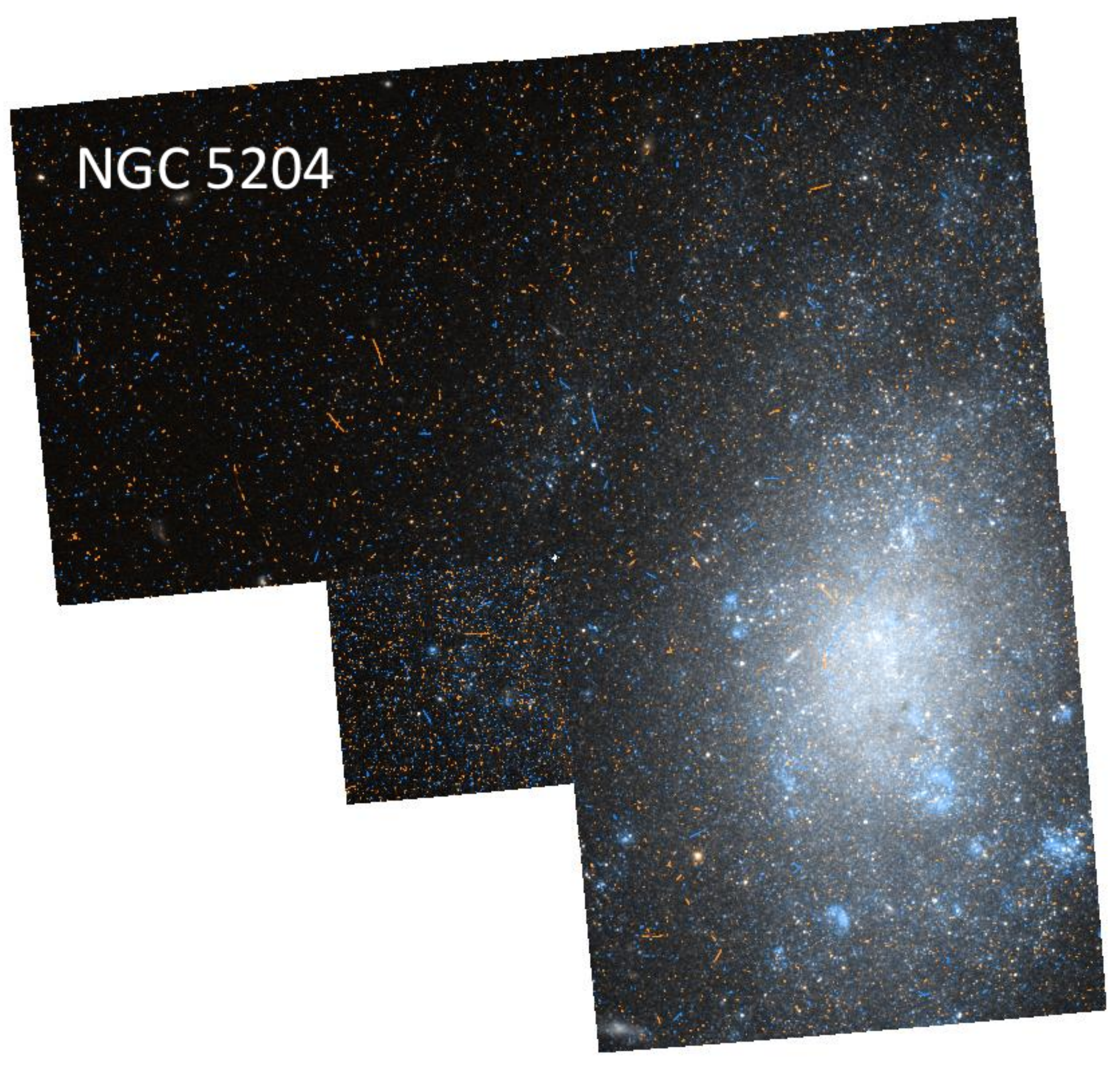}
\caption{Composite color HST images of NGC~4190 and NGC~5204 showing both the main star forming complexes and underlying older stellar populations extending to greater radii. The images are from the Hubble Legacy Archive (Image credit: HLA, ESA, NASA) and were created by combining the F606W data (blue) and the F814W data (red). The fields are oriented North-up and East-left.}
\label{fig:images}
\end{figure*}

\section{Observations and Photometry of the Resolved Stellar Populations\label{obs}}
The data for both NGC~4190 and NGC~5204 consist of archival HST imaging obtained using the Wide Field Planetary Camera 2 (WFPC2) with the F606W and F814W filters. Total integration times per filter were 2200 s for NGC~4190 and 600 s for NGC~5204. Table~\ref{tab:properties} summarizes the observations and basic properties of the systems. Figure~\ref{fig:images} shows composite color images of both galaxies produced from the Hubble Legacy Archive pipeline. Visible in both galaxies are blue, presumably young, clustered, star-forming regions as well as a more homogeneously distributed redder, presumably older stellar population. 

\begin{table}
\begin{center}
\caption{Galaxy Properties and Observations}
\label{tab:properties}
\vspace{-20pt}
\end{center}
\begin{tabular}{lrr}
\hline 
\hline 
  					& NGC~4190			& NGC~5204 		\\
\hline
R.A. (J2000) 			& 12:13:44.8			& 13:29:36.5	 	\\
Decl. (J2000)			& $+$36:38:02.5  		& $+$58:25:07.4	\\
Program ID			& HST-GO-10905		& HST-GO-8601	\\
WFPC~2 Filters		& F606W; F814W		& F606W; F814W	\\
Exposure per filter (s)	& 2200				& 600			\\
Distance (Mpc)			& 2.83$\pm0.17$		& 4.65$\pm0.53$	\\
Best-fit Distance (Mpc)	& 2.95$\pm0.07$		& 4.79$\pm0.11$	\\
Foreground A$_R$ (mag)	& 0.063				& 0.027			\\
Best-fit A$_{F606W}$ (mag)& $0.00\pm0.05$	& $0.20\pm0.05$\\
\hline
\end{tabular}
\textbf{Notes.} The distances were measured using the tip of the red giant branch (TRGB) method \citep{Tully2009, Jacobs2009, Karachentsev2003}. Galactic extinction is based on the \citet{Schlafly2011} recalibration of the \citet{Schlegel1998} dust maps. The best-fit distance and extinction values are from the CMD-fitting technique discussed in the text.
\end{table}

To ensure a uniform comparison of the star-formation properties with the starburst sample from \citet{McQuinn2010b}, we processed the data and measured the SFHs with the same methodology and using the same tools as this previous study. The images were cosmic-ray cleaned and processed using the standard HST pipeline. Point spread function (PSF) photometry was performed on the individual images using HSTphot \citep{Dolphin2000} which is optimized for the under-sampled PSF in WFPC2 data. The raw photometry lists were filtered for well-recovered stellar-like point sources with a signal-to-noise ratio $\geq4$ and a data flag $\leq1$. The resulting photometric catalog was further filtered for point sources with low sharpness and crowding values: $|$sharpness values$| \leq 0.3$ and crowding values $\leq 0.5$. The sharpness parameter aids in rejecting extended, background sources. The crowding parameter aids in removing stars whose photometric measurements are significantly affected in crowded regions by neighboring point sources. Artificial stars tests were performed on the images to measure the completeness limit and noise characteristics of the data, using the same photometry package and filtered with the same parameters.

\begin{figure*}
\includegraphics[width=0.49\textwidth]{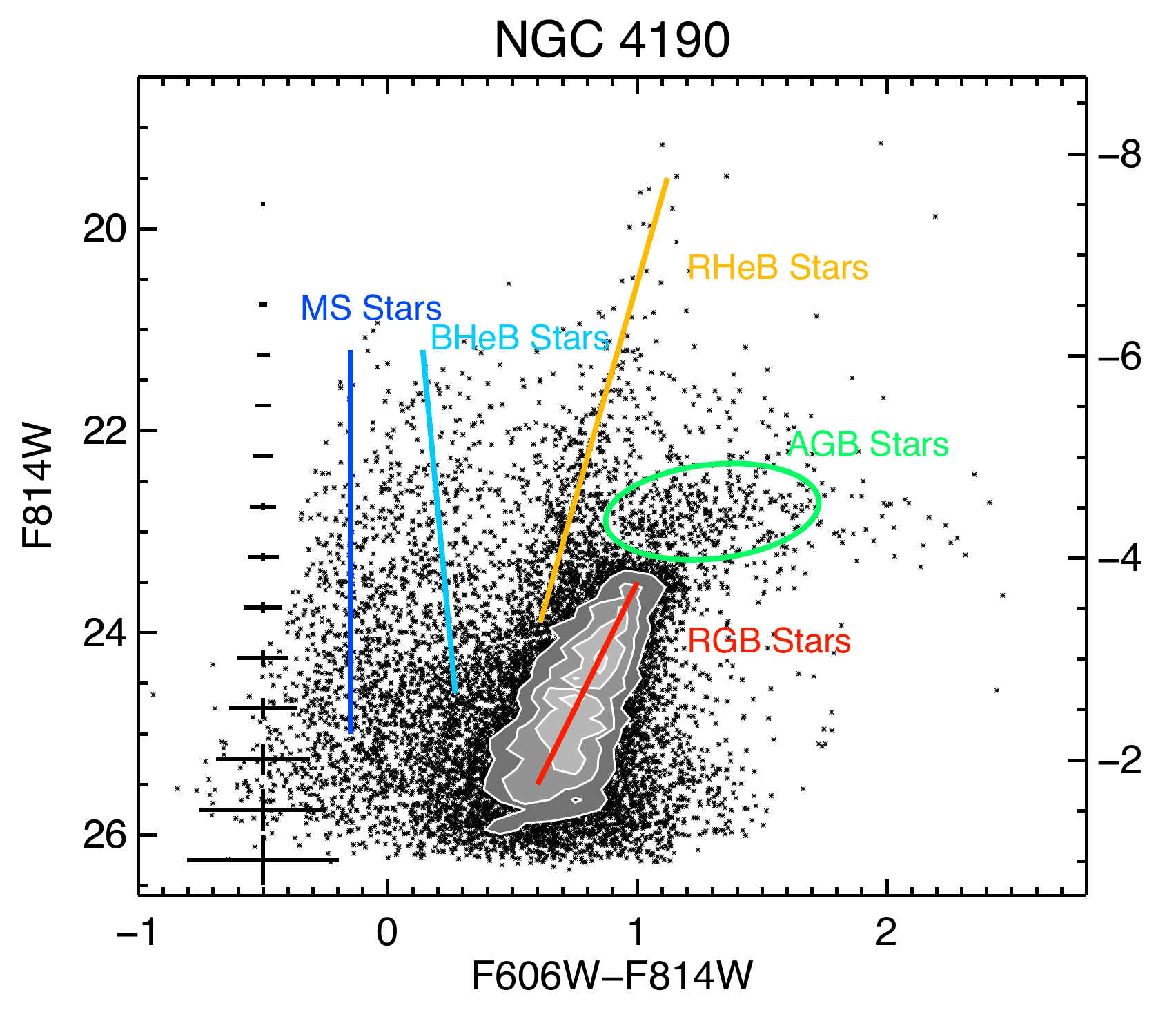}
\includegraphics[width=0.49\textwidth]{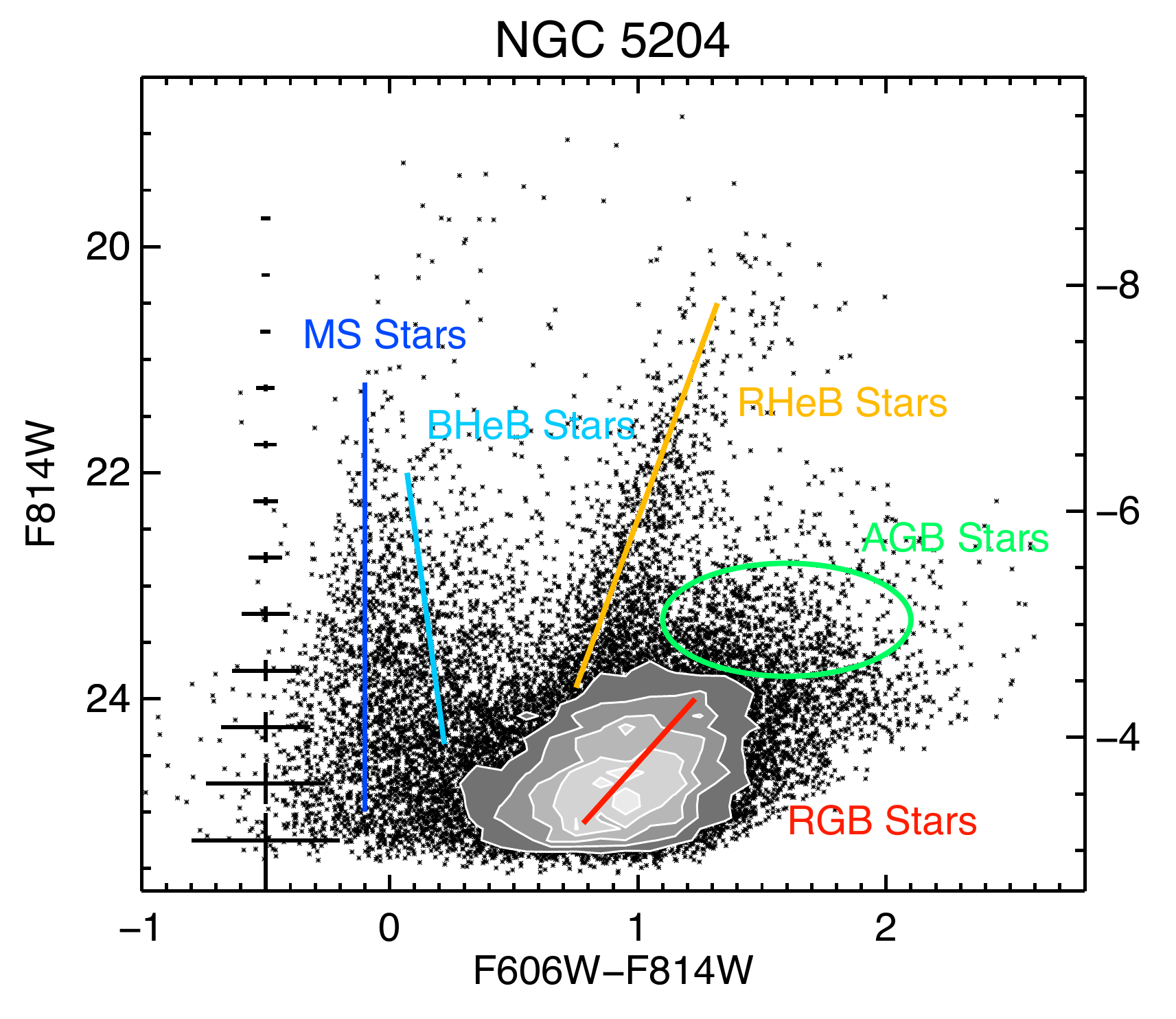}
\caption{CMDs of NGC~4190 and NGC~5204. The RGB, MS, AGB, and red and blue HeB sequences are well-populated in both CMDs. The abundance of intermediate mass HeB stars is a clear sign of recent star formation in these galaxies.}
\label{fig:cmds}
\end{figure*}

Figure~\ref{fig:cmds} shows the CMDs for both galaxies plotted to the 50\% completeness level as determined from the artificial star tests. Apparent in each CMD are well-defined red giant branch (RGB) sequences populated by stars older than $\sim$1 Gyr, upper main sequence (MS) stars, asymptotic giant branch (AGB) stars, and both red and blue helium burning (HeB) sequences. Stars on the HeB sequences are of intermediate mass (M$\gtrsim3$\msun) and are burning helium in their cores \citep{Bertelli1994}. These stars are relatively young, with ages ranging from 5 to 1000 Myr. During this time period, the stars will evolve off the MS to the red HeB sequence, migrate back across the CMD to the blue HeB sequence, and finally evolve into an AGB star or end as a supernova. In contrast to RGB and MS stars, the location of an HeB star on the CMD depends uniquely on age with more massive, younger HeB stars being more luminous than less massive, older HeB stars. Observationally, HeB stars can be used as chronometers of star formation in a galaxy over the last $\ltsimeq500$ Myr, when the HeB sequences typically merge into the red clump and become indistinguishable from other stars \citep{Dohm-Palmer1997, Dohm-Palmer2002, McQuinn2011}. 

\section{The Star Formation Histories}
The SFHs of both galaxies were measured using the numerical CMD fitting program {\tt MATCH} \citep{Dolphin2002}. This is the same SFH code used in the starburst study of \citet{McQuinn2010a} ensuring a uniform comparison of starburst metrics. Briefly, for an assumed initial mass function (IMF), {\tt MATCH} uses stellar evolutionary isochrones to create a series of synthetic stellar populations of different ages and metallicities. The modeled CMD that is the best-fit to the observed CMD provides the most likely SFH of the galaxy. The synthetic CMDs are modeled using the photometry and recovered fractions of the artificial stars as primary inputs. The Padua stellar evolution models were used \citep{Marigo2008} including the updated AGB tracks from \citet{Girardi2010}. We assumed a Salpeter IMF \citep{Salpeter1955} from 0.1 to 120 \msun\ and a binary fraction of 35\% with a flat secondary distribution. 

Distance and extinction are free parameters fit by {\tt MATCH}. The best-fit distances are reported in Table~\ref{tab:properties}. They agree within the errors with independent distance measurements based on the tip of the red giant method, providing a consistency check on our results. Both foreground and internal extinction can broaden the features in a CMD, but they are expected to be low for these high-latitude, metal-poor galaxies \citep[cf.,][]{Berg2012}. Table~\ref{tab:properties} lists both the foreground extinction estimate from the dust maps of \citet{Schlegel1998} with recalibration by \citet{Schlafly2011} and the best-fit extinction values, which includes both foreground and internal extinction at a slightly shorter wavelength. Because the photometric depth of the data does not constrain the metallicity evolution of the systems, we assumed that the chemical enrichment history, $Z(t)$, is a continuous, non-decreasing function over the lifetime of the galaxy. Uncertainties on the SFHs take into account both systematic uncertainties from the stellar evolution models \citep{Dolphin2012} and random uncertainties due to the finite number of stars in a CMD \citep{Dolphin2013}. 

Measuring a SFH requires sufficient information from stellar populations of different ages. At older ages, the best temporal resolution achievable is largely a function of photometric depth \citep{Dolphin2002, McQuinn2010a}. Photometry reaching the oldest MS turn-off is required to distinguish between stars that are, for example, 12 Gyr or 10 Gyr old. Thus, with the exception of the closest galaxies in the Local Group, the temporal resolution achievable at the earliest epochs is limited to multi-Gyr bins. At younger ages, the best temporal resolution achievable is largely a function of the number of stars in the upper part of the CMD \citep{Dolphin2013} with a lesser dependency on photometric depth. Thus, in galaxies with recent star formation, both the upper MS and HeB sequences can be used to constrain the recent SFHs with a much higher temporal resolution than is achievable at older times (see Section~2). The random uncertainties measured for the SFHs can help in choosing the appropriate temporal resolution for an individual data set; if the time binning is too fine, the uncertainties increase substantially. We chose the time binning for the SFHs balancing the desire for the maximum time resolution with the size of the random uncertainties. This is an improvement over the approach used in \citet{McQuinn2010a}, where only systematic uncertainties were included in the SFH and a uniform time binning scheme was adopted based on the photometric depth of the data. This difference in temporal resolution will not impact our measurement of whether or not these two galaxies meet the starburst criteria because our time binning is well within the typical durations of starbursts in low-mass galaxies \citep[a few 100 Myr;][]{McQuinn2010b}.

\section{Are These Galaxies Starburst Dwarfs?}
\begin{figure*}
\includegraphics[width=\textwidth]{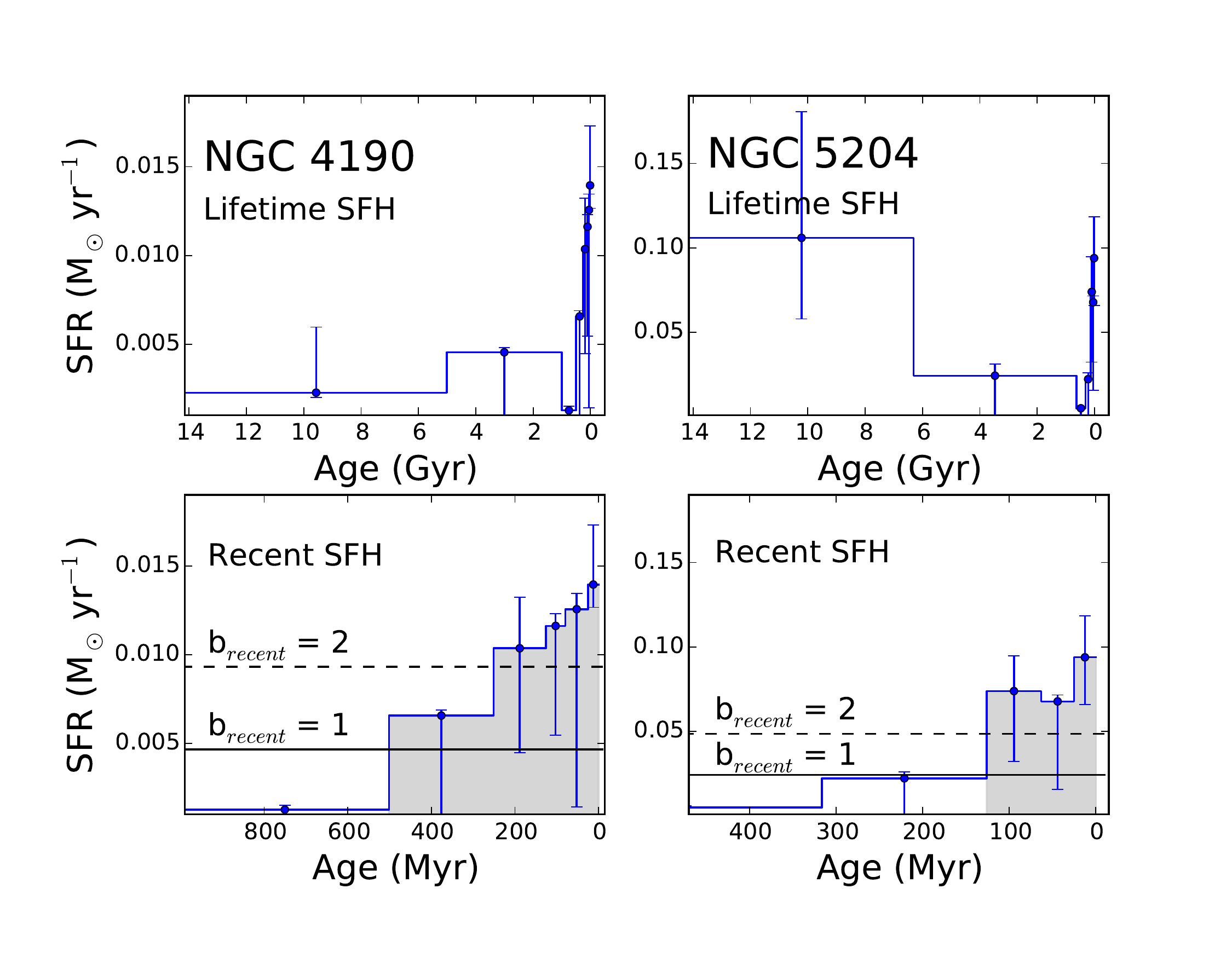}
\vspace{-40pt}
\caption{The lifetime SFHs of NGC~4190 and NGC~5204 are plotted in the top panels; an expanded view of the most recent  SFHs are plotted in the bottom panels. Both galaxies show an increase in star-formation activity at recent times. The solid lines mark $b_{recent} = 1$, while the dotted lines mark $b_{recent} = 2$. According to the criteria of \citet{McQuinn2010b}, both galaxies host a starburst with a duration of $380\pm130$ Myr in NGC 4190 and $100\pm30$ in NGC 5204 (shaded gray regions).}
\label{fig:sfhs}
\end{figure*}

In Figure~\ref{fig:sfhs} we present the lifetime and recent SFHs of NGC~4190 and NGC~5204. Despite differences in the earliest epochs of star formation, both NGC~4190 and NGC~5204 show a significant rise in SFRs at recent times compared with the average SFR over the last few Gyr. This rise in recent star-formation activity is very similar to those reported by \citet{McQuinn2010b} in a sample of galaxies comprised of known starbursts.

Multiple studies have used comparisons of recent to past star-formation activity to identify starbursts, generally setting a birthrate parameter threshold of $2-3$ to separate burst from non-burst activity \citep[e.g.,][]{Hunter1986, Salzer1989, Gallagher2005, Kennicutt2005, Lee2009}. Many $b$ measurements use recent star-formation activity measured over very short timescales of $5-10$ Myr corresponding to the lifetimes of the most massive stars (traced by H$\alpha$ emission) and are susceptible to biases from short-duration stochastic fluctuations in individual star-forming regions. Such ``flickering'' star-formation activity can be 10$\times$ greater than SFRs measured over longer time periods \citep{Lee2009}. The lifetime SFR averages are often based on integrated spectral measurements which can be highly uncertain \citep{Kennicutt1994}. The value of $b$ has a strong redshift dependence with peak values as high as 30 for a wide demographic of starburst galaxies \citep{Brinchmann2004}. In studies of a narrower demographic of galaxies in the nearby universe, the average birthrate parameter based on disc-averaged SFRs for non-burst galaxies is $\sim0.5$, thus a $b$ threshold value of 2 requires recent star-formation activity to be $4\times$ greater in starbursts than in typical non-bursting galaxies \citep[e.g.,][]{Kennicutt2005}. 

\citet{McQuinn2010b} followed a similar convention to identify and measure the characteristics of starburst in the SFHs, but modified the birthrate parameter to focus on the more recent evolutionary state of the host galaxies. Specifically, the birthrate parameter was defined as $b_{recent} \equiv \rm{SFR/\overline{SFR}}_{t<6~Gyr}$, where the normalization factor is the historical average SFR within 6 Gyr. The SFRs at the earliest epochs (t $> 6$ Gyr) are not included in the historical averages, because (i) the uncertainties are larger at these older times, (ii) the generally higher levels of star-formation activity during the epoch of galaxy assembly can bias the averages, and (iii) the ``recent'' evolutionary state of the galaxy (t $\leq 6$ Gyr) is a more relevant baseline with which to compare the SFRs at the present epoch. This approach is only possible when measuring SFHs as other star-formation indicators are restricted to using rather uncertain lifetime average SFRs as a birthrate normalization factor. Following the convention that $b_{recent} \geq 2$ identifies a starburst, \citet{McQuinn2010b} further prescribed that birthrate values $\simeq1$ can be used to identify the beginning and end of a burst. Using this metric, the duration of a starburst is measured as the amount of time $b_{recent} \geq1$, given the $b_{recent} \geq 2$ threshold has been reached. 

As seen in Figure~\ref{fig:sfhs}, both NGC~4190 and NGC~5204 have SFRs at recent times that are higher than the $b_{recent}$ threshold of 2 and, thus, can be considered starbursts. Moreover, the disc-averaged recent SFRs for both galaxies are elevated for at least 100 Myr, and continue to be higher than the past averages at the present day. The durations of the starbursts, shaded in grey, are $>380\pm130$ Myr for NGC~4190 and $>100\pm30$ Myr for NGC~5204.  The peak $b_{recent}$ values are $3.0^{+0.7}_{-0.6}$ and $3.9^{+1.0}_{-1.4}$ for NGC~4190 and NGC~5204 respectively (see Table~\ref{tab:sf_characteristics}). In both cases the $\rm{\overline{SFR}}_{t<6~Gyr}$ normalization factor for the b parameters are upper limits, thus the $b_{recent}$ values may be higher. Fluctuations in the star-formation activity and higher birthrate values over shorter time periods are presumably present in the galaxies, but the ability to measure such fluctuations is limited by the time resolution achievable with the current data. Despite this limitation, the $b_{recent}$ values are still above the starburst threshold over multiple time bins, reinforcing the conclusion that these are starbursts rather than stochastic fluctuations in the SFRs corresponding to individual pockets of star formation. Similar to the starburst sample from \citet{McQuinn2010b}, the measured durations are longer than or equivalent to the dynamical timescales of the host galaxies calculated from gas rotation speeds and effective radii (i.e., $\tau_{dyn} \equiv 2\pi R_{eff} / V_{rot}$; see Table~\ref{tab:sf_characteristics}). Additionally, in Table~\ref{tab:sf_characteristics} we provide quantities calculated from the SFH characterizing the recent star formation including the percent of mass created in the bursts and the energy outputs of the bursts from supernovae and stellar winds estimated using the Starburst99 model \citep{Leitherer1999}. These values also fall within the ranges measured for the starburst sample in \citet{McQuinn2010b}.

In summary, the star-formation characteristics in NGC~4190 and NGC~5204 are consistent with the properties measured in a larger sample of known starburst galaxies from \citet{McQuinn2010b}. Indeed, these star-forming events are comprised of multiple star-forming regions with elevated levels of star formation lasting 100 Myr or longer, rather than being localized in individual, short-duration pockets of star formation.

\begin{table}
\begin{center}
\caption{Measured Star-Formation Characteristics}
\label{tab:sf_characteristics}
\end{center}
\vspace{-20pt}
\begin{tabular}{lrr}
\hline 
\hline 
  								& NGC~4190					& NGC~5204 			\\
\hline
Peak SFR of Burst (\msun\ yr$^{-1}$)	& 14$^{+3}_{-1}\times10^{-3}$		& 9$^{+2}_{-3}\times10^{-2}$	\\
Peak $b_{recent}$ of Burst			& $3.0^{+0.7}_{-0.6}$			& $3.9^{+1.0}_{-1.4}$	\\
Duration (Myr)						& $\geq380\pm130$				& $\geq100\pm30$		\\
Total Stellar Mass (\msun)				& 4.4$^{+3.5}_{-2.5}\times10^7$	& 9.8$^{+6.3}_{-5.2}\times10^8$	\\
Percent Mass created in burst			& 8$^{+6}$\%					& 0.7$^{+0.5}$\%		\\
log Energy from Burst (erg)			& 55.9						& 56.2				\\
R$_{eff}$ (kpc)						& 3.8	$\pm0.2$					& 11.3$\pm1.3$			\\
V$_{rot}$ (km s$^{-1}$)				& 25$\pm3$					& 63$\pm4$			\\
$\tau_{dyn}$ (Myr)					& 93$\pm12$					& 110$\pm14$			\\
Duration/$\tau_{dyn}$				& 4.1$\pm1.5$					& 0.9$\pm0.3$			\\
\hline
 \\
\end{tabular}
\textbf{Notes.} The measured star-formation characteristics of NGC~4190 and NGC~5204. Both galaxies meet the criteria of starburst from \citet{McQuinn2010b}, have starburst durations lasting 100 Myr or more, and star formation characteristics consistent with known starbursts. The stellar mass listed is the total mass formed in each galaxy (not the current stellar mass); the percent of mass formed in the bursts are upper limits. Effective radii are from \citet{Swaters2002b}; rotational velocities are from \citet{Swaters2009}.
\end{table}

\section{Discussion \& Conclusions}
In this study we measured the SFHs of two compact dwarf galaxies (NGC~4190 and NGC~5204) that were selected based solely on their proximity and dynamical properties: \hi rotation curves with a steep rise in the inner regions ($V_{R_{\rm d}}/R_{\rm d} \gtrsim 50$ km~s$^{-1}$~kpc$^{-1}$). We found that the star-formation properties of both galaxies are consistent with those of previously studied, known starbursts \citep{McQuinn2010b}. This strongly reinforces the idea that the starburst activity is closely linked to the central distribution of mass (gas, stars, and dark matter), as suggested by \citet{Lelli2012a, Lelli2012b, Lelli2014a}. 

Understanding the link between starbursts and mass distribution is challenging as it depends on distinguishing between \textit{nature} versus \textit{nurture} scenarios: either the progenitors of starburst dwarfs are unusually compact dIrrs with steeply rising rotation curves, or there must be a mechanism that concentrates the mass in a typical dIrrs and eventually triggers the starburst (or some combination of both). 

The former possibility (\textit{nature}) has been discussed in detail by \citet{Meurer1998}, \citet{vanZee2001}, and \citet{Lelli2014a}. According to the Toomre instability criteria \citep{Toomre1964}, a galaxy with a steeply rising rotation curve in the inner parts has a high value of the critical surface-density threshold for large-scale gravitational instabilities. As a consequence, the gas may pile up in the center over a cosmic time and reach high surface densities, until the critical threshold is reached and a starburst develops. As pointed out by \citet{Lelli2014a}, this does not necessarily imply a SFH characterized by short bursts and long quiescent periods, since the star formation may continue in the outer regions of the disc (along the flat part of the rotation curve) where the disc may be less stable.

The latter possibility (\textit{nurture}) requires a mechanism (internal or external) that can transform a typical dwarf into a compact one. Several internal mechanisms have been proposed, such as mass inflows due to clump instabilities \citep{Elmegreen2012}, triaxial dark matter haloes \citep{Bekki2002}, or bars made of dark matter \citep{Hunter2004}. These mechanisms are generally difficult to test with observations; to date there is no direct evidence that they may actually take place in dwarf galaxies. External mechanisms, instead, can be investigated using deep optical imaging \citep[e.g.,][]{LopezSanchez2010b, Delgado2012} or \hi observations \citep[e.g.,][]{Ekta2008, Ekta2010a}. Recently, \citet{Lelli2014c} studied the distribution and kinematics of the \hi gas in the \textit{outer regions} of 18 starburst dwarfs, and found that they systematically have more asymmetric/disturbed \hi morphologies than typical dIrrs. As discussed in \citet{Lelli2014c}, these outer \hi asymmetries cannot be the result of massive gas outflows \citep[see also][]{Lelli2014b}, but point to external mechanisms triggering the starbursts. Obvious possibilities are interaction/mergers between dwarf galaxies or cold gas accretion from the intergalactic medium. NGC~4190 and NGC~5204 qualitatively fit this picture: the \hi map of NGC~4190 shows pronounced asymmetries in the outer parts, while NGC~5204 has a strongly warped \hi disc (see Appendix~B of \citealt{Swaters2002a}). Interestingly, numerical simulations suggest that the formation of a compact mass distribution may be easily explained by interactions/mergers between gas-rich dwarfs \citep[e.g.,][]{Bekki2008}, but a detailed comparison between simulations and observations is still lacking.

Regardless of the \textit{nature} versus \textit{nurture} issue, it would be interesting to search for the descendants of starburst dwarfs, or post-starburst dwarfs, among quiescent compact dwarfs. If the link between mass concentration and starbursts is confirmed, it would also have important implications for statistical estimates of burst duty cycles based on large galaxy surveys \citep[e.g.,][]{Lee2009}, given that only compact dIrrs would repeatedly experience starburst episodes during their lifetime. Moreover, it would bring into question whether stellar feedback from starbursts can really transform central dark matter cusps (predicted by N-body simulations) into the observed cores \citep[e.g.,][]{Governato2010, Oh2011b, Governato2012, Teyssier2013, Garrison-Kimmel2013, DiCintio2014, Brooks2014, Ogiya2014, Madau2014}.  SFHs of other compact dIrrs from new HST imaging would clarify whether dIrrs with steeply-rising rotation curves are necessarily experiencing a starburst (like NGC~4190 and NGC~5204) or whether some of them can be considered descendants of starburst dwarfs. This would shed new light on the future evolution of starburst dwarfs and firmly establish the link between star formation and mass distribution for a large galaxy sample.

\section*{Acknowledgments}
FL thanks Filippo Fraternali, Renzo Sancisi, and Marc Verheijen for many enlightening discussions about the dynamics of dwarf galaxies. This research made use of NASA's Astrophysical Data System and the NASA/IPAC Extragalactic Database (NED) which is operated by the Jet Propulsion Laboratory, California Institute of Technology, under contract with the National Aeronautics and Space Administration.

\renewcommand\bibname{{References}}
\bibliographystyle{mn2e.bst}
\bibliography{../bibliography.bib}

\end{document}